\documentclass[%
 reprint,
 amsmath,amssymb,
 aps,
]{revtex4-1}

\makeatletter
\newcommand*{\rom}[1]{\expandafter\@slowromancap\romannumeral #1@}
\makeatother

\usepackage{siunitx}
\usepackage[caption=false]{subfig}
\usepackage{graphicx}
\usepackage{dcolumn}
\usepackage{bm}

\begin{document}

\preprint{APS/123-QED}

\title{Importance of Electronic Correlation in the Intermetallic Half-Heusler Compounds}

\author{Minjie Lu}
 \affiliation{Physics Department, Texas A\&M University.}
\author{Hao Chen}
 \affiliation{Physics Department, Texas A\&M University.}

\author{Glenn Agnolet}
 \email{agnolet@tamu.edu}
 \affiliation{Physics Department, Texas A\&M University.}

\date{\today}

\begin{abstract}
Low temperature scanning tunneling spectroscopy of HfNiSn shows a $V^m(m < 1)$ zero bias anomaly around the Fermi level. This local density of states with a fractional power law shape is well known to be a consequence of electronic correlations. For comparison, we have also measured the tunneling conductances of other half-Heusler compounds with 18 valence electrons. ZrNiPb shows a metal-like local density of states, whereas ZrCoSb and NbFeSb show a linear and $V^{2}$ anomaly. One interpretation of these anomalies is that a correlation gap is opening in these compounds. By analyzing the magnetoresistance of HfNiSn, we demonstrate that at low temperatures, electron-electron scattering dominates. The $T^m(m<1)$ temperature dependence of the conductivity confirms that the electronic correlations are a bulk rather than a surface property. 
\begin{description}
\item[PACS numbers]
May be entered using the \verb+\pacs{#1}+ command.
\end{description}
\end{abstract}

\pacs{Valid PACS appear here}
\maketitle


\section{\label{sec:level1}INTRODUCTION}

Recently, there has been renewed interest in more efficient thermoelectric materials because of the worldwide demand for sustainable energy sources. One such class of candidates are the half-Heusler compounds with 18 valence electrons. Among them, the MNiSn (M = Ti, Zr or Hf) compounds are narrow gap semiconductors with a gap size of 100-\SI{300}{\meV} \cite{Cui01, Han03}. The narrow gap leads to a moderate electrical resistivity and a large Seebeck coefficient. This combination leads to a high $ZT$, the figure of merit used to evaluate thermoelectric efficiency \cite{Cui01, Shu02, Han03}. $ZT = \alpha^2T/(\rho\kappa)$, where $\alpha$ is the Seebeck coefficient, $\rho$ is the electronic resistivity, $\kappa$ is the thermal conductivity and $T$ is the absolute temperature. Intensive efforts have been made to improve the ZT of MNiSn compounds. Because MNiSn compounds are n-type materials, all of the three sites, M, Ni and Sn, can be doped by donors to reduce the electronic resistivity \cite{64, 65, 66}. Grain size and grain boundaries can also be engineered to achieve a better Seebeck coefficient \cite{63, 79}. However, these attempts can be facilitated by a better understanding of the electronic structure near the Fermi level.

Half-Heusler compounds have the MgAgAs-type crystal structure which belongs to the F-43m space group. The crystal structure is shown in Fig. \ref{fig:images}(a). It is comprised of four interpenetrating face-centered cubic sublattices. Two sublattices are occupied by transition metals. The third is occupied by a main group element and the fourth is occupied by vacancies. The vacancy sublattice distinguishes half-Heusler compounds from their full-Heusler counterparts as it induces a gap in the electronic structure. The intrinsic disorder observed in many half-Heusler compounds \cite{7, 11, Han03} is known to be donors or acceptors that can generate in-gap states \cite{11, 12, 235}. Annealing can reduce the concentration of this disorder and transform the compounds from a metal to an insulator \cite{7}. Close to the metal-to-insulator transition, disorder breaks down momentum conservation and electronic screening becomes inefficient. Therefore, a strong electronic correlation is expected.

In this paper, we present scanning tunneling spectroscopy (STS) of four half-Heusler compounds. Except for ZrNiPb which has a metal-like local density of states (LDOS), the other compounds have different power law $V^{\alpha}$ zero bias anomalies (ZBA) near the Fermi level, around which narrow but hard gaps should be expected. These anomalies are characteristic LDOS of a disorder induced metal-to-insulator transition (MIT) \cite{Mcmillan18, 238}. Details are discussed in Section \rom{3}.A. These giant anomalies in the one-electron local density of states play an important role in determining the electronic and thermal properties of these compounds. Ab-initio approaches generally fail to predict their existence \cite{Colinet08, 11} and therefore, the subsequent conclusions from these approaches are unreliable. These anomalies also make it difficult to interpret the results of doping experiments because, instead of only shifting the chemical potential, doping also leads to the evolution of the anomaly \cite{201, 202}. In Section \rom{3}.B and \rom{3}.C, the magnetoresistance and the temperature dependence of conductivity of HfNiSn are analyzed to provide more evidence for a strong electronic correlation in the material.

\section{EXPERIMENTAL DETAILS}
The HfNiSn samples are flux-grown single crystals provided by our collaborator Dr. Lucia Steinke \cite{23}, who also provided the resistivity and magnetoresistance data. Details of sample growth and measurement of properties are described in \cite{23}. The ZrNiPb, ZrCoSb and NbFeSb samples are polycrystals provided by our collaborator Dr. Fei Tian of Dr. Zhifeng Ren's group \cite{64}. Tunneling conductances are measured with a low temperature scanning tunneling microscope (STM) located in a \textsuperscript{3}He cryostat. Atomic resolution images of graphite and self-assembled dodecanethiol monolayer on an atomically flat gold surface have been obtained with this STM. Mechanically cut Platinum-Iridium tips are used. In order to get reproducible tunneling conductance and because of a lack of in-situ tip treatment device, tips are annealed in a Bunsen flame \cite{Libioulle22}. Using the same kind of tip, no ZBA is observed in the tunneling conductance of gold down to \SI{77}{\kelvin}. Samples are cleaved right before experiments to expose fresh surfaces. Differential conductances are measured by a standard lock-in technique with a 0.5 mV modulation.

\section{RESULTS AND DISCUSSION}

\subsection{\label{sec:level2}Local density of states}

\begin{figure*}

\subfloat[]{%
	\includegraphics[clip,width=0.3\paperwidth]{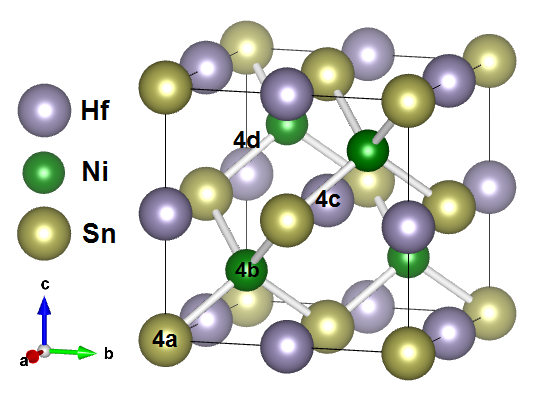}%
}
\subfloat[]{%
  	\includegraphics[clip,width=0.23\paperwidth]{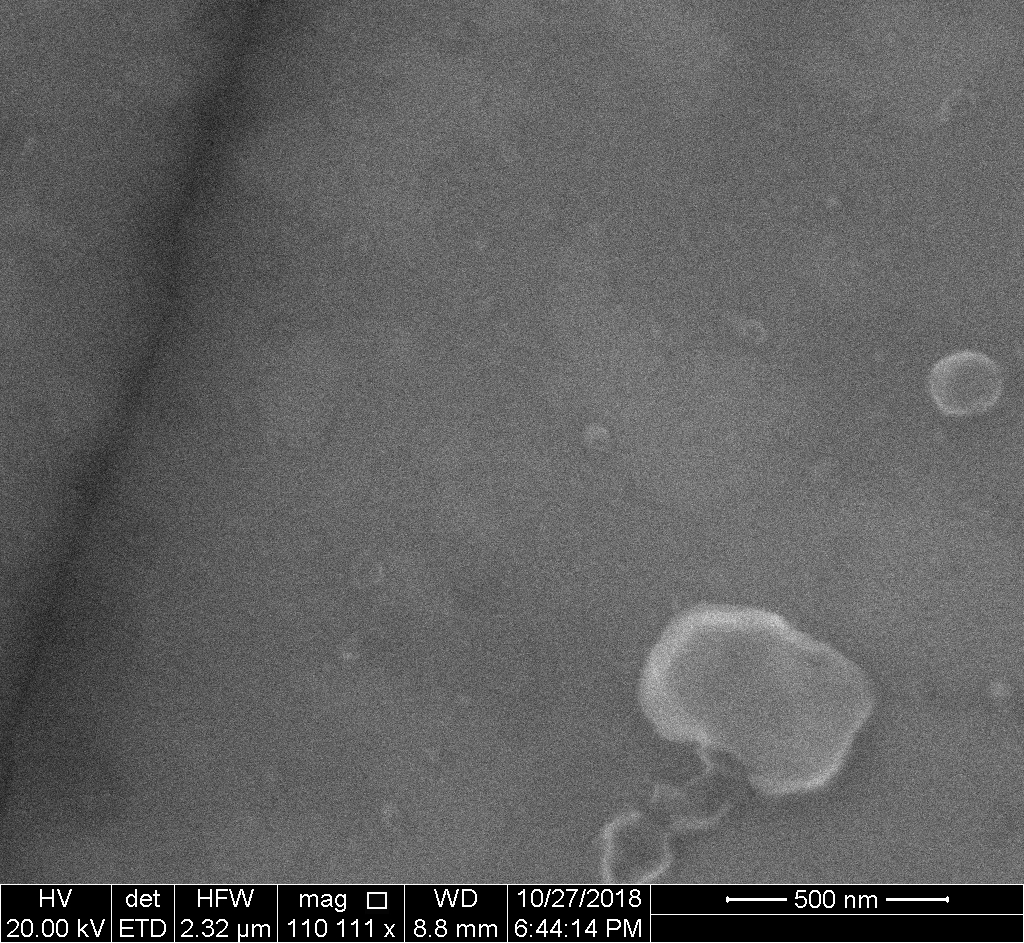}%
}
\subfloat[]{%
  	\includegraphics[clip,width=0.212\paperwidth]{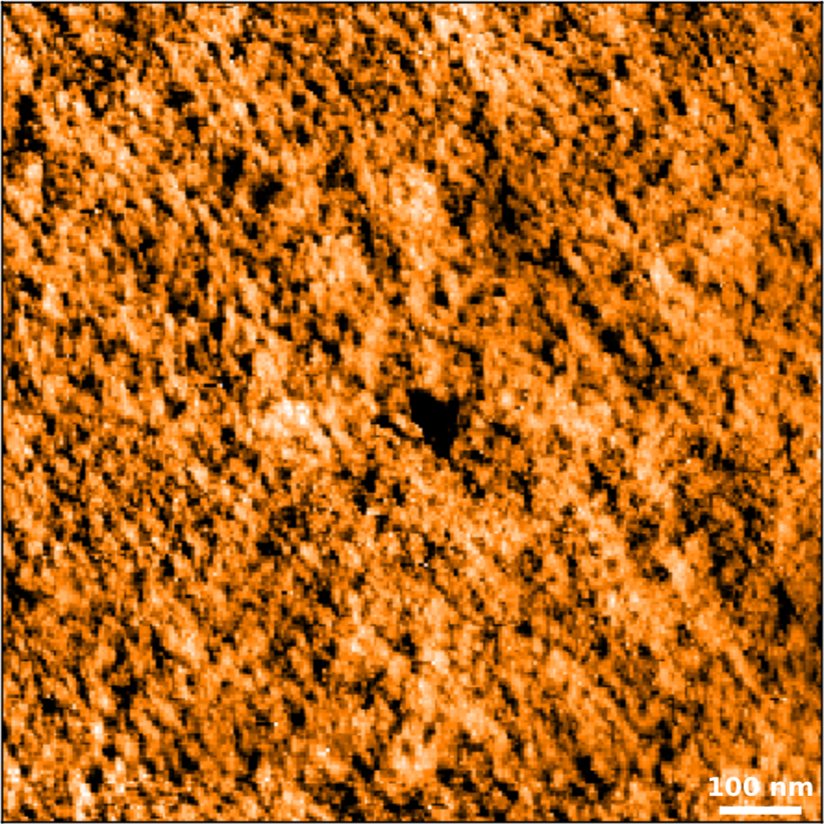}%
}

\caption{\label{fig:images}(a) Crystal structure of HfNiSn. 4 Wyckoff positions are labelled. The 4d position is occupied by vacancies. (b) SEM image of a freshly cleaved surface. (c) STM image of the same surface.}

\end{figure*}

Figure. \ref{fig:images}(b, c) show a comparison of a scanning electron microscope (SEM) and an STM image of a freshly cleaved HfNiSn surface. Although the STM reveals structures beyond the resolution of the SEM, the rough surface makes it difficult to achieve atomic resolution on these materials. Such a cratered surface is expected because of the vacancy lattice and the intrinsic disorder.

\begin{figure}

\subfloat[]{%
	\includegraphics[clip,height = 0.185\paperheight, width=0.6\columnwidth]{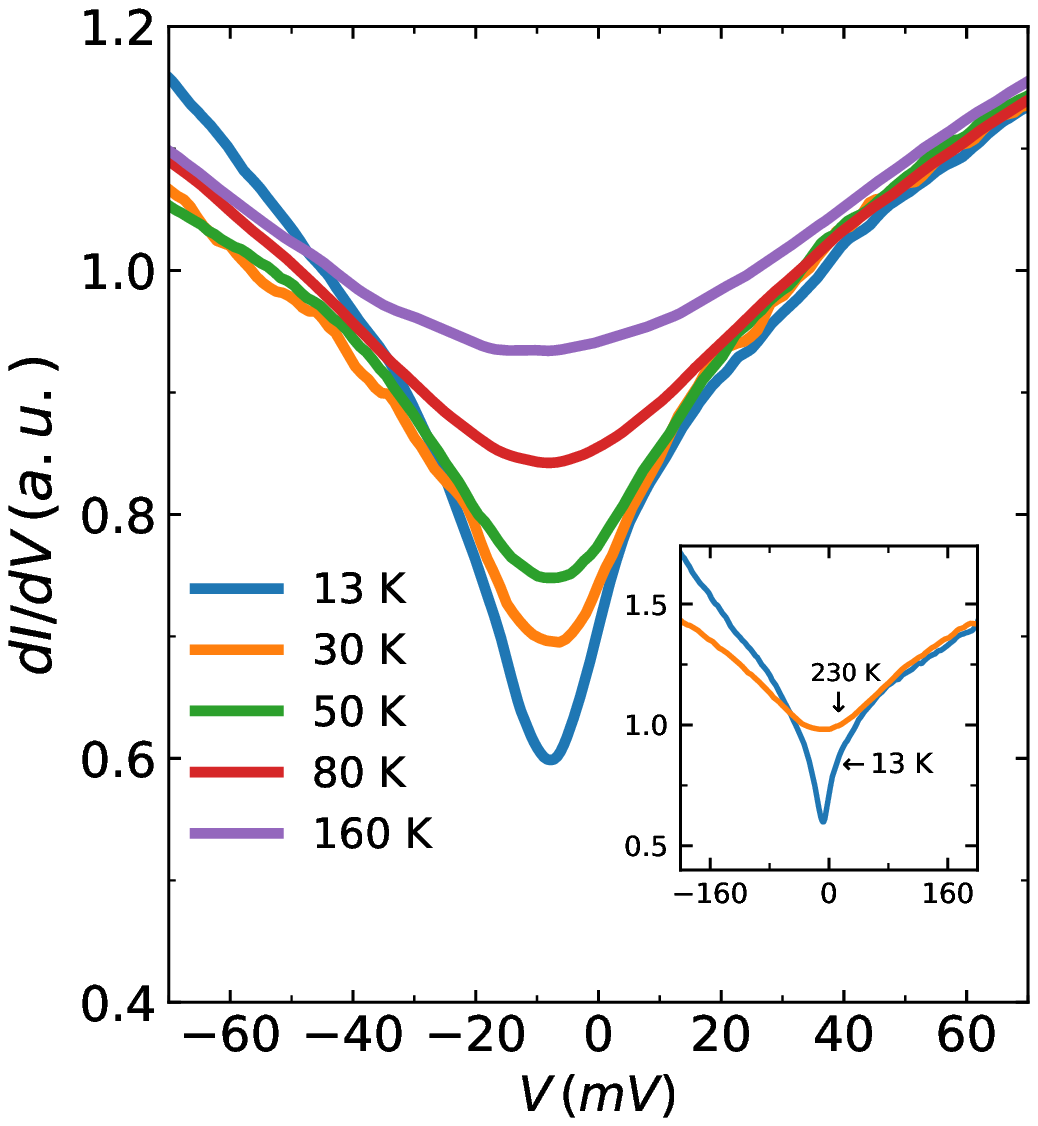}
	\vspace{1cm}
}
\subfloat[]{%
  	\includegraphics[clip,height = 0.18\paperheight, width=0.4\columnwidth]{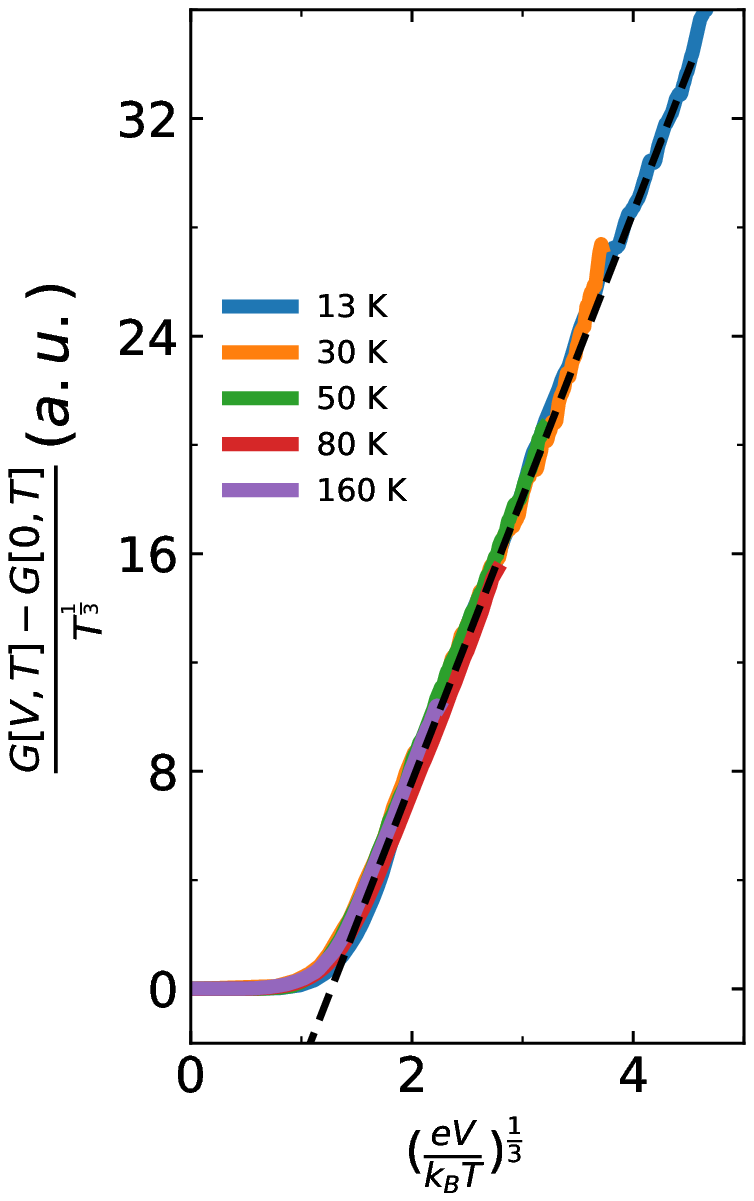}%
}

\caption{\label{fig:ldos}(a) Normalized tunneling conductances of a single crystal HfNiSn from 13 K to 160 K. The different data sets were normalized by overlapping their values at high positive bias. Inset shows tunneling conductances at \SI{13}{ \kelvin} and \SI{290}{\kelvin} over a larger bias range. (b) Plot of $[G(V,T)-G(0,T)]/T^{\frac{1}{3}}$ versus $(eV/k_BT)^{\frac{1}{3}}$. All curves between 13 K and 160 K collapse to a universal one. The black dashed line is a linear fit indicating a $V^{\frac{1}{3}}$ power law dependence.}

\end{figure}

The tunneling conductances of two HfNiSn samples are more systematically studied. One shows a $V^{\frac{1}{2}}$ ZBA, common in disordered conductors with electronic correlation. The other shows a $V^{\frac{1}{3}}$ ZBA. The tunneling conductances of the latter at different temperatures are plotted in Fig. \ref{fig:ldos}(a). All the minima occur at \SI{-7}{mV} instead of at the Fermi level. The inset shows the conductances measured at \SI{13}{\kelvin} and \SI{290}{\kelvin} over a larger bias range, from which the $V^{\frac{1}{3}}$ shape is clear. Although the tunneling conductance does vary slightly across the surface, the dependence on voltage remains the same. Figure \ref{fig:ldos}(b) plots $(G[V,T]-G[0,T])/T^{\frac{1}{3}}$ versus $(eV/k_BT)^{\frac{1}{3}}$, where $G[V,T]$ is the tunneling conductance, and $V$ is the bias voltage. The straight dashed line corresponds to a $V^{\frac{1}{3}}$ dependence. Curves at different temperatures collapse to a universal one that is expected from the finite temperature density of states (DOS) calculated by Altshuler and Aronov \cite{39}.

Tunneling conductance of nominal semiconductors can be sensitive to surface defects. But the $\sim T^m$ conductivity discussed in Section \rom{3}.C supports that it is the bulk rather than surface electrons that are correlated. For films thinner than tens of nanometers, the conductivity should be proportional to $ln(T)$ \cite{15, 39}, whereas the corrugation of the rough HfNiSn surface is around \SI{2}{\nm} from the STM topographic image. The $\sqrt{B}$ dependence of magnetoresistance at high fields also demonstrates that the transport is 3 dimensional \cite{242}.

\begin{figure}

\subfloat[]{%
	\includegraphics[clip, height = 0.17\paperheight, width=0.5\columnwidth]{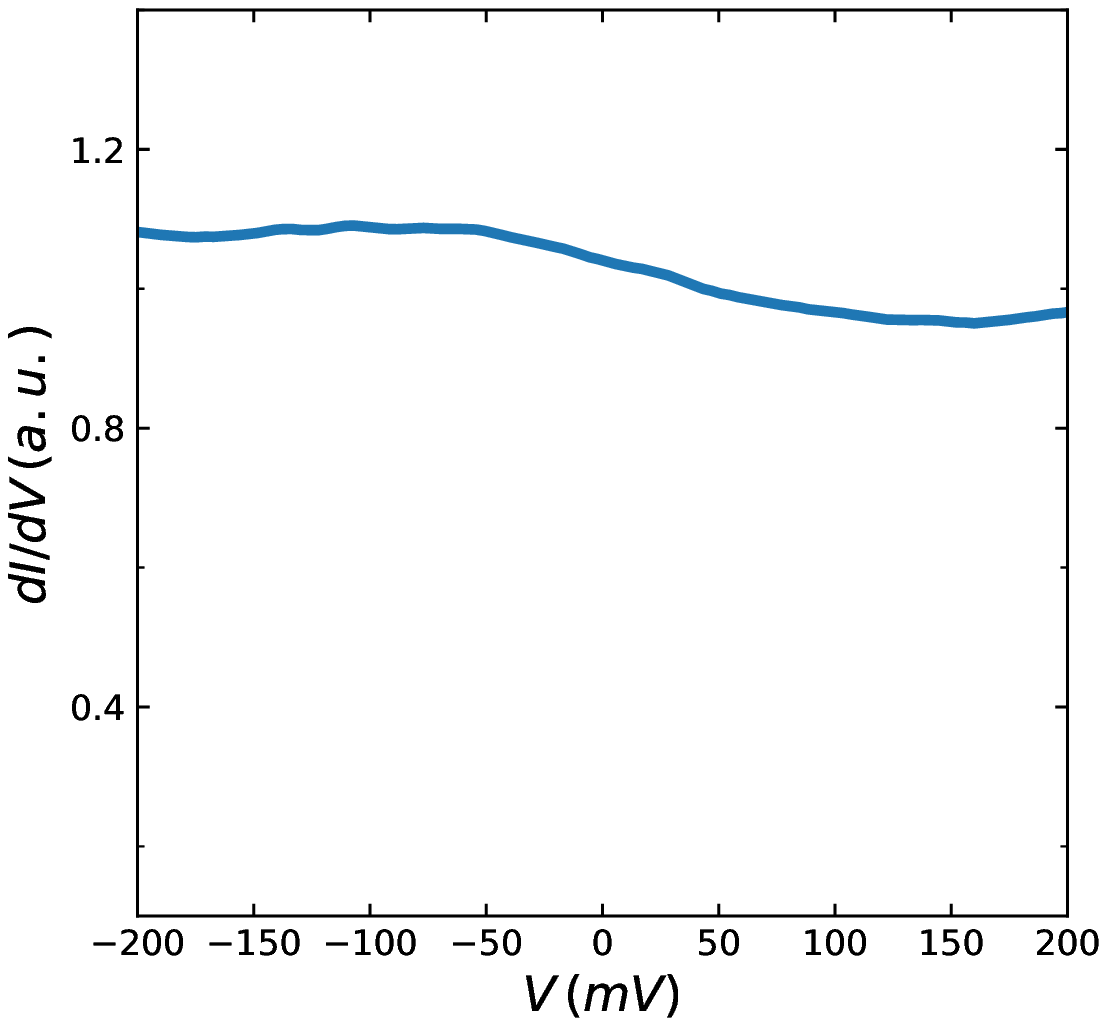}%
}
\subfloat[]{%
	\includegraphics[clip, height = 0.17\paperheight, width=0.5\columnwidth]{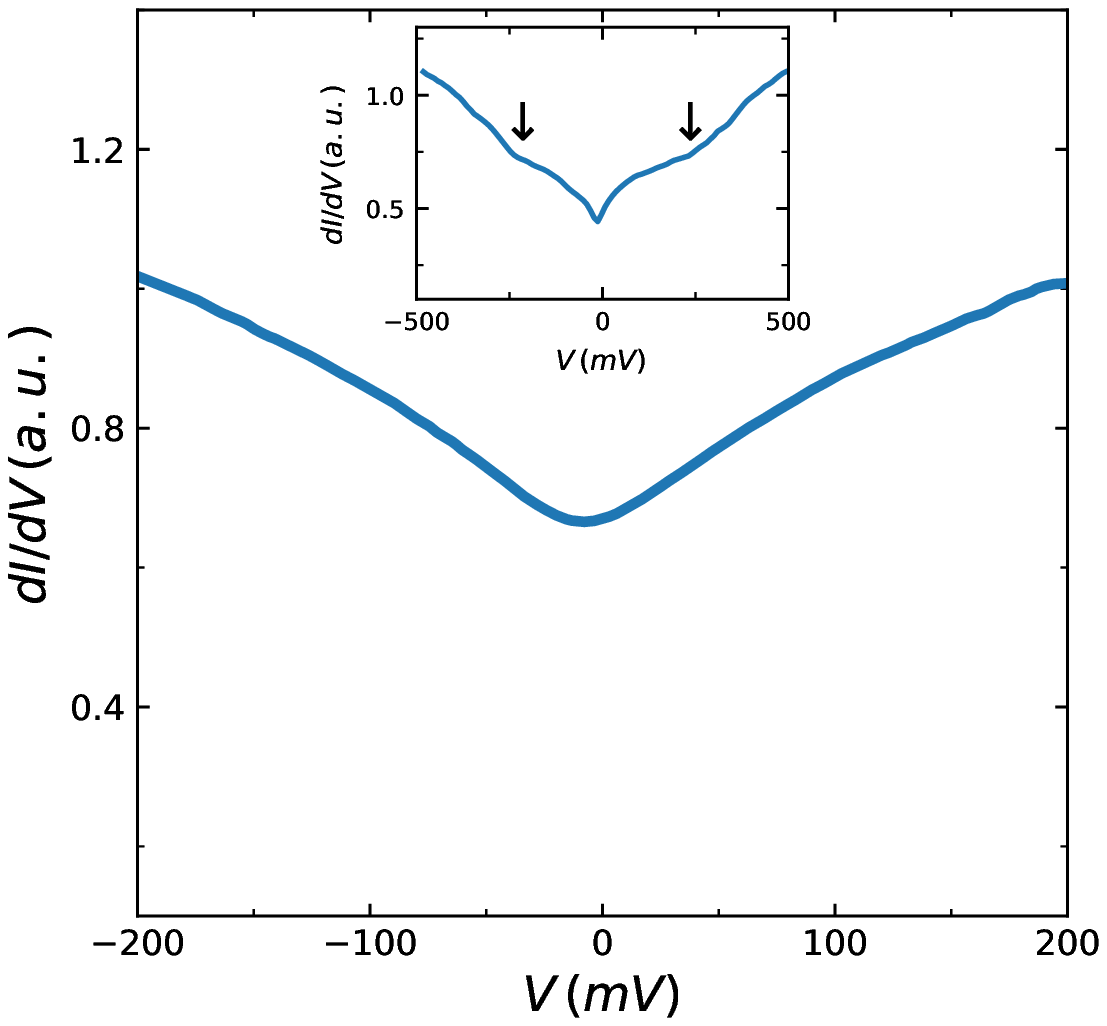}%
}
\\
\subfloat[]{%
  	\includegraphics[clip, height = 0.17\paperheight, width=0.5\columnwidth]{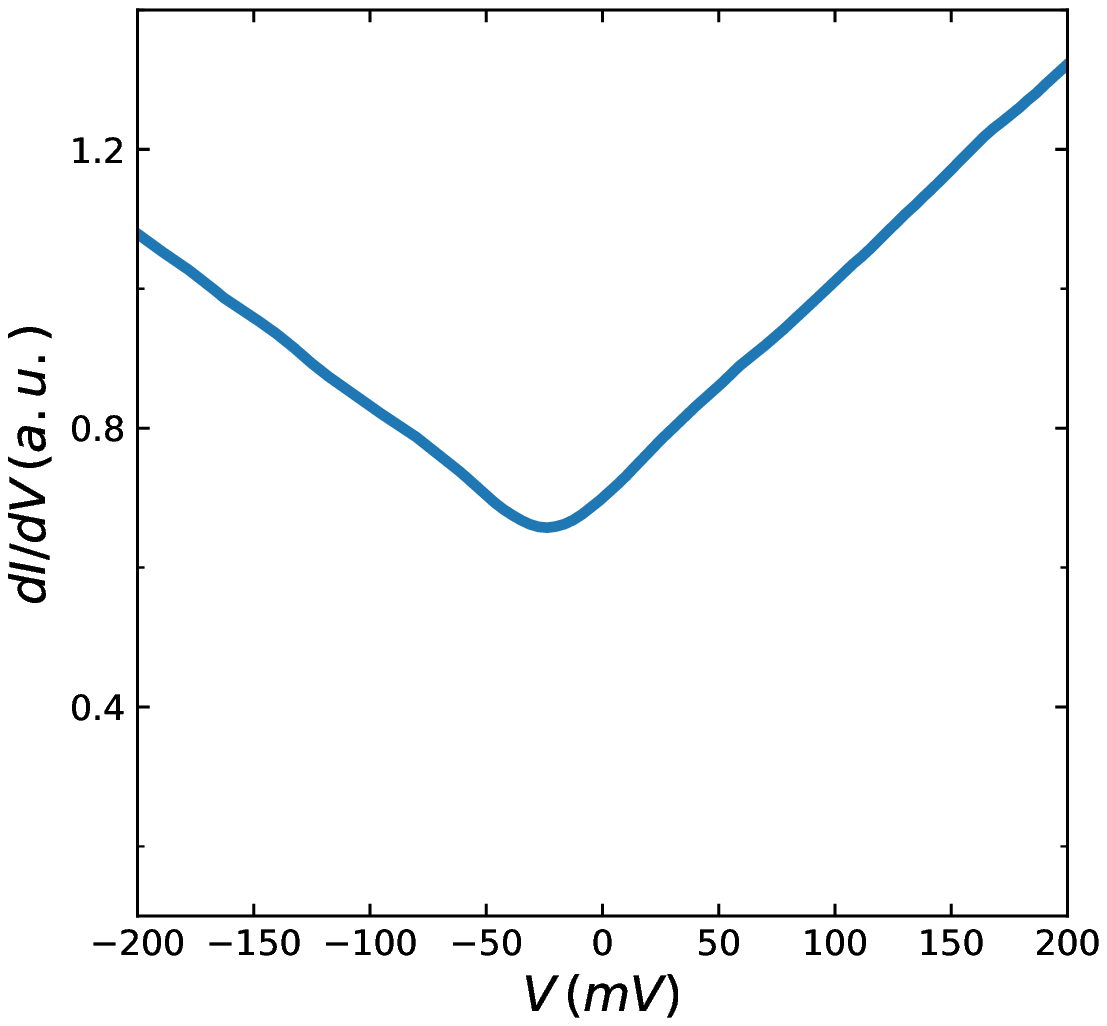}%
}
\subfloat[]{%
	\includegraphics[clip, height = 0.17\paperheight, width=0.5\columnwidth]{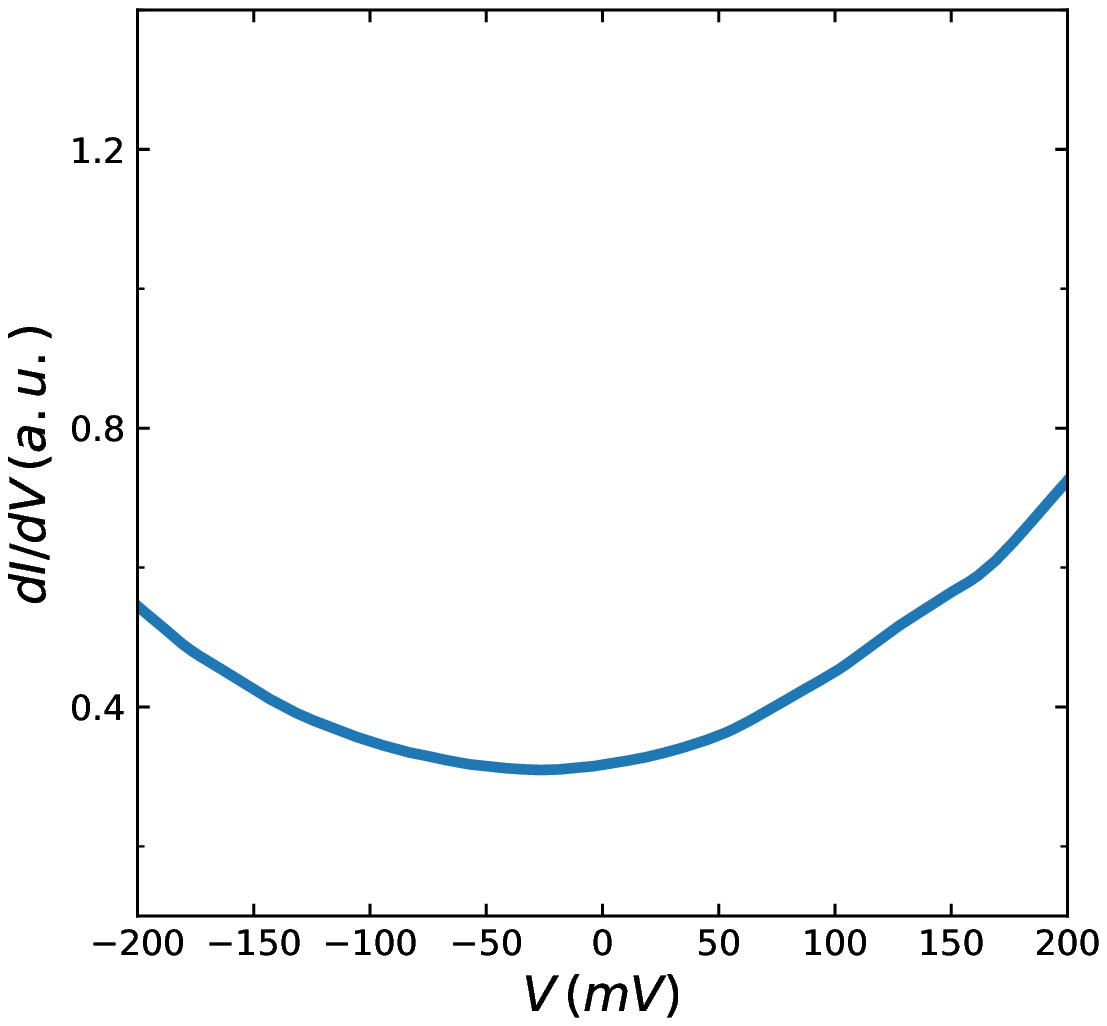}%
}

\caption{\label{fig:differentmaterials} Normalized tunneling conductance of (a) ZrNiPb, (b) HfNiSn, (c) ZrCoSb and (d) NbFeSb. The inset of (b) shows the the tunneling conductance of HfNiSn from \SI{-500}{\meV} to \SI{500}{\meV}. Arrows point to kinks which possibly indicate band edges. The inset of (d) shows the corresponding current-voltage curve of NbFeSb.}

\end{figure}

We further measured the tunneling conductances of other half-Heusler compounds, ZrNiPb, ZrCoSb and ZrCoSb. These samples are polycrystals. ZrNiPb and HfNiSn are n-type semiconductors \cite{107, 212}, whereas ZrCoSb and NbFeSb are p-type \cite{64, 212}. The results are shown in Fig. \ref{fig:differentmaterials}. ZrNiPb shows a metal-like LDOS, whereas ZrCoSb and NbFbSb show a linear and a quadratic ZBA separately. The quadratic ZBA is likely the correlation gap broadened by both thermal smearing and interactions \cite{217, 236}. These measurements are taken at room temperature simply to observe the giant ZBA indicating electronic correlation. Low temperature measurements may reveal other interesting features for detailed discussion. For each compound, we reproduced these observations on at least two samples indicating that these tunneling conductances are the characteristic of the different compounds. Except for the different electronic structures of the pure compounds, their chemical bonds may favor different types and concentrations of the intrinsic disorder.

\subsection{\label{sec:level2}Magnetoresistance}

In this section, we analyze the magnetoresistance of HfNiSn to provide another evidence of strong electronic correlation.

\begin{figure}

\subfloat[]{%
	\includegraphics[clip, height = 0.2\paperheight, width=0.85\columnwidth]{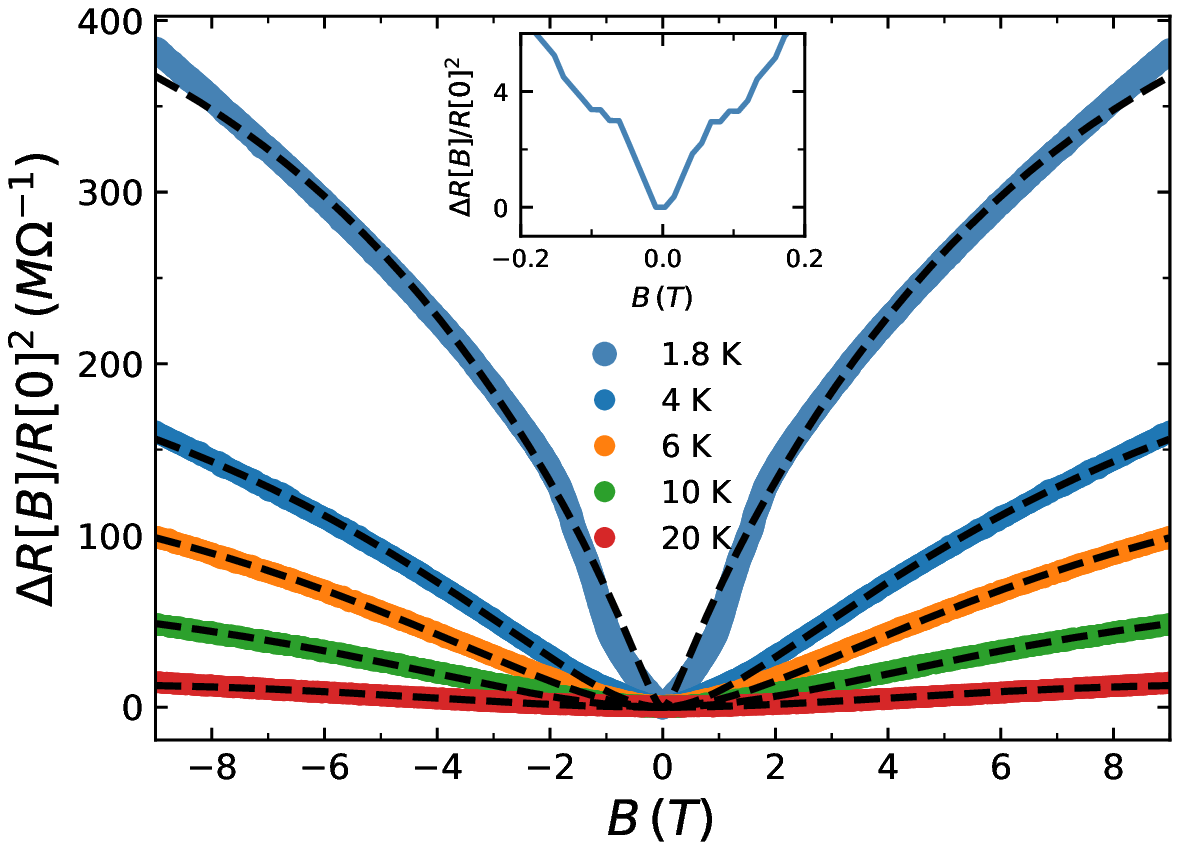}%
}
\\
\subfloat[]{%
  	\includegraphics[clip, height = 0.201\paperheight, width=0.42\columnwidth]{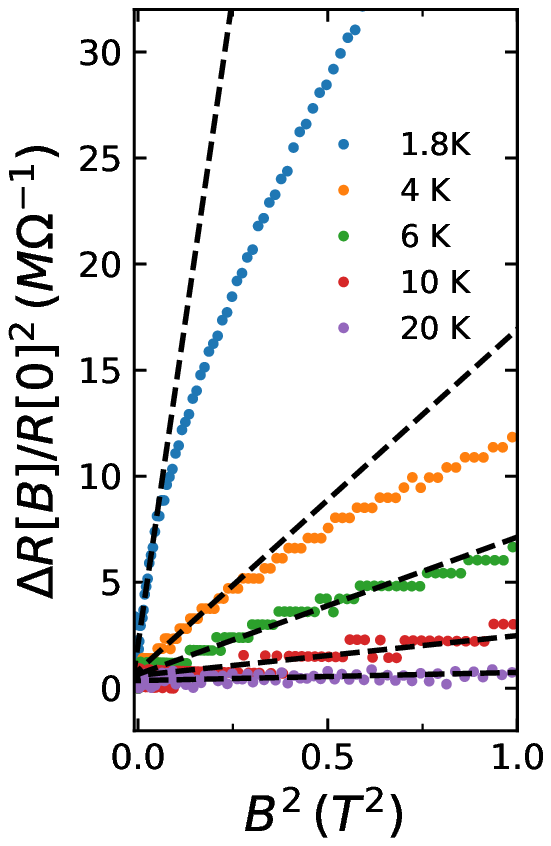}%
}
\quad
\subfloat[]{%
	\includegraphics[clip, height = 0.2\paperheight, width=0.42\columnwidth]{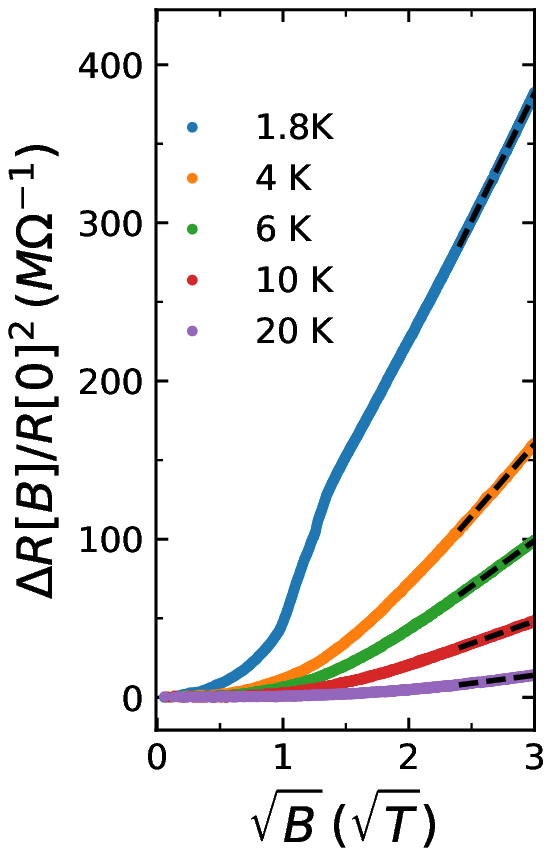}%
}

\caption{\label{fig:magnetoresistance}(a) Magnetoresistance $\Delta R[B]/R[0]^2=(R[B]-R[B=0])/R[B=0]^2$ of a single crystal HfNiSn at different temperatures. The black dashed lines are the best fits generated using a sum of the F-H formula and the corrections due to electronic correlation. (b) Magnetoresistance versus $B^2$. Dashed lines are linear fits. (c) Magnetoresistance versus $\sqrt{B}$. Data are provided by our colleague. \cite{23}}

\end{figure}

The magnetoresistance of HfNiSn at different temperatures are shown in Fig. \ref{fig:magnetoresistance}(a). The $B^2$ dependence at low field and the $\sqrt{B}$ or a weaker dependence at high fields are indicators of weak anti-localization (WAL) \cite{52, 53, 57}. WAL is a quantum phenomenon observed in disordered electronic systems caused by the interference of self-crossing trajectories of backscattered electrons. With a strong spin-orbit coupling \cite{55} or a $\pi$ Berry's phase \cite{26, 27}, the interference is destructive and the current is enhanced. Magnetic field adds additional phases to the electrons, thereby destroying the interference causing an increase in the resistance. 

The formula for 3D WAL is provided by Fukuyama and Hoshino (F-H) \cite{31, 32}. The correction due to electronic correlation consists of two parts. The orbital part is calculated by Altshuler \textit{et al.} \cite{226} and the spin part is calculated by Lee and Ramakrishnan \cite{227}. Alternative expressions for the orbital part are also available \cite{39, 228}. These formula are suspected to fail at high fields \cite{229, 230} and many fits to experimental data show clear deviations \cite{51, 55, 56}. Therefore, we generate the dashed fitted lines in Fig. \ref{fig:magnetoresistance}(a) merely to show that the magnetoresistance of HfNiSn is a typical WAL magnetoresistance, but the fitting parameters will not be discussed here.

WAL has become an important tool to extract the electron dephasing scattering times $\tau_{\phi}$ \cite{51, 55, 56, 230}. The critic field $B_{\phi} = B_i + 2B_s$ contained in the F-H formula is inversely proportional to $\tau_{\phi}$. $B_x = \frac{\hbar}{4eD\tau_x}$, where $x$ stands for ($i$) inelastic, ($s$) spin-flipping or ($so$) spin-orbit scattering respectively, and D is the temperature independent diffusion coefficient related to elastic scattering \cite{31, 230, 231}. Because of the aforementioned reason, instead of fitting the magnetoresistance, we adopt another approach to get the temperature dependence of $\tau_\phi$.

At low field and low temperature, $B \ll B_{\phi} \ll B_{so}$. Utilizing the low field analytic expression of the $f_3$ function provided by Ousset \textit{et al.} \cite{232}, the F-H formula reduces to a very simple form.

\begin{equation}
\frac{\Delta R[B]}{R[0]^2} = \frac{c_1}{B_{\phi}^\frac{3}{2}}B^2
\label{eqs:lowfieldfh}
\end{equation}

\noindent where $\Delta R[B] = R[B] - R[0]$, $c_1 = \frac{ce}{192\pi^2} (\frac{e}{\hbar})^\frac{3}{2}$ and $c$ is a constant related to the geometry of the sample.

$B_{\phi}$ provides information on different scattering mechanisms. $B_{\phi} = B_i + 2B_s \sim \frac{1}{\tau_i} + \frac{2}{\tau_s}$, where $\tau_s$ is a temperature independent scattering time due to magnetic impurities. Usually, $\frac{1}{\tau_i} = \frac{1}{\tau_{ee}} + \frac{1}{\tau_{ep}}$, where $\tau_{ee}$ is the electron-electron scattering time and $\tau_{ep}$ is the electron-phonon scattering time. It has been established that $\tau_{ep} \sim T^{-x}$, where $x$ can be 2, 3 or 4 \cite{56, 207}. In disordered systems, at low temperature, $\tau_{ee} \sim T^{-\frac{d}{2}}$, where $d$ is the dimension of the electronic transport in the material \cite{39, 56, 59}.

At low fields, utilizing the analytic expression of the $\Phi_3$ function provided by Ousset \textit{et al.} \cite{232}, the orbital part of the correction due to electronic correlation becomes

\begin{equation}
\frac{\Delta R[B]}{R[0]^2} = \frac{c_2(c_3 - lnT)}{T^{\frac{3}{2}}}B^2
\label{eqs:lowfieldspin}
\end{equation}

\noindent where $c_2 = 0.32925 \times \frac{ce}{2\pi^2}(\frac{2De^2}{\pi k_B\hbar})^{\frac{3}{2}}$, $c_3 = \frac{1}{\lambda} + ln\frac{\gamma T_F}{\pi}$, $\lambda$ is the temperature independent electron-phonon coupling constant \cite{233, 234}, $\gamma = 0.577$ is Euler's constant and $T_F$ is a cut off temperature. 

Utilizing the analytic expression of the $g_3$ function provided by Ousset \textit{et al.} \cite{232}, the spin part becomes

\begin{equation}
\frac{\Delta R[B]}{R[0]^2} = \frac{c_4}{T^{\frac{3}{2}}}B^2
\label{eqs:lowfieldorbital}
\end{equation}

\noindent Where $c_4 = 0.056464\times \frac{c(eg\mu_B)^2F}{8\pi^3\sqrt{2D}(\hbar k_B)^{\frac{3}{2}}}$, $g = 2$ is the electron spin g-factor, $\mu_B$ is the Bohr magneton, $F(0<F<1)$ is the screening parameter for the Coulomb interaction.

\begin{figure}
\centering
\includegraphics[width=67mm, height = 60mm]{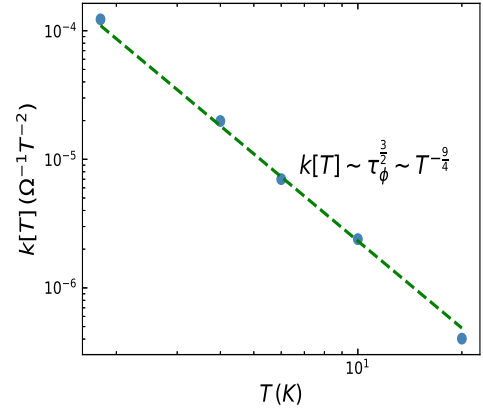}
\caption{\label{fig:fittingparameters}Log-log plot of $k[T]$ versus temperature.}
\end{figure}

The data clearly follows a $B^2$ dependence at low fields. Taken together, Eqs. \ref{eqs:lowfieldfh}, \ref{eqs:lowfieldspin} and \ref{eqs:lowfieldorbital} indicate that $\frac{\Delta R[B]}{R[0]^2} = k[T]B^2$ where the slopes of the dashed lines in Fig. \ref{fig:magnetoresistance}(b) indicates the values of $k[T]$. The determined values of $k[T]$ are very robust. For example, in the analysis of the \SI{4}{\kelvin} magnetoresistance, regressions are performed using increasing numbers of data points from $B^2 < \SI{0.1}{\tesla}^2$ to $B^2 < \SI{0.2}{\tesla}^2$. The calculated $k[T]$ slowly decreases from $1.99\times10^{-5}$ to $1.74\times10^{-5}$ which will not qualitatively change the conclusion. Figure \ref{fig:fittingparameters} is the log-log plot of $k[T]$ versus $T$. The almost linear relation represents either the $T^{-\frac{3}{2}}$ term in Eq. \ref{eqs:lowfieldspin} and \ref{eqs:lowfieldorbital} or one of the scattering mechanisms in $B_{\phi}$ dominates. By minimizing the mean squared error of $log(k[T])$, the green dashed line in Fig. \ref{fig:fittingparameters} is generated using only Eq. \ref{eqs:lowfieldfh} and supposing that $B_{\phi} \sim T^{\frac{3}{2}}$, when only the 3D electron-electron scattering is considered. One could also attempt to fit to an expression that includes Eq. \ref{eqs:lowfieldfh}, \ref{eqs:lowfieldspin}, \ref{eqs:lowfieldorbital} and all possible $T^x$ terms in $B_{\phi}$. However, when the $T^{-\frac{3}{2}}lnT$ term in Eq. \ref{eqs:lowfieldspin} is excluded, no better fitting can be generated than simply including the effect of electron-electron scatterings. Therefore, the magnetoresistance supports a strong electronic correlation and a 3D electronic transport in HfNiSn.

\subsection{\label{sec:level2}Conductivity}

According to McMillan \cite{14}, Altshuler and Aronov \cite{39}, electronic correlation leads to a quantum correction to the the Boltzmann conductivity in weakly disordered systems, $\sigma_I = \sigma_0 + AT^{\frac{1}{2}}$. This relation correctly describes the typical low temperature conductivity of HfNiSn from \SI{1}{\kelvin} to around \SI{140}{\kelvin}.

High temperature conductivity of HfNiSn implies a gap of \SI{100}{\meV} to \SI{300}{\meV} \cite{7, 34, 70}. Our flux-grown HfNiSn single crystals also show an activated conductivity above \SI{200}{\kelvin}. But the calculated gap is around \SI{490}{\meV}. In the tunneling conductance, there is a small upturn at $\mid V \mid = \SI{250}{\mV}$ shown in the inset of Fig. \ref{fig:differentmaterials}(b). If the upturn indicates the band edge, the tunneling conductance also indicates a comparable gap of \SI{500}{\meV} and the ZBA probably lies in an impurity band. 

\section{Conclusion}

In conclusion, we observed ZBA of various power laws in the STS of ZrNiPb, HfNiSn and NbFeSb. These ZBA, that are probably in the impurity band, are the result of disorder and strong electronic correlations. Within the framework of traditional ab initio methods, it is difficult to explain these ZBA and consequent properties of half-Heusler compounds with 18 valence electrons. Similarly, the discussions of doping experiments of these compounds should be reexamined. From the magnetoresistance of HfNiSn, we found that below \SI{20}{\kelvin}, the inelastic scattering time $\tau_{in}$ is proportional to $T^{-\frac{3}{2}}$, which indicates that electron-electron scattering dominates. We also showed that the low temperature conductivity of HfNiSn obeys $\sigma = \sigma_0 + AT^{\frac{1}{2}}$, also indicating a strong electronic correlation.

\nocite{apsrev41Control}
\bibliography{HfNiSn}

\end{document}